\documentstyle[12pt]{article}
\input{psfig}
\def\la{\mathrel{\mathpalette\fun <}}
\def\ga{\mathrel{\mathpalette\fun >}}
\def\fun#1#2{\lower3.6pt\vbox{\baselineskip0pt\lineskip.9pt
  \ialign{$\mathsurround=0pt#1\hfil##\hfil$\crcr#2\crcr\sim\crcr}}}
\def\mathrelfun#1#2{\lower3.6pt\vbox{\baselineskip0pt\lineskip.9pt
  \ialign{$\mathsurround=0pt#1\hfil##\hfil$\crcr#2\crcr\sim\crcr}}}

\def\eV  {{\rm \hbox{e\kern-0.14em V}}}
\def\keV {{\rm \hbox{ke\kern-0.14em V}}}
\def\MeV {{\rm \hbox{Me\kern-0.14em V}}}
\def\GeV {{\rm \hbox{Ge\kern-0.14em V}}}

\input epsf.tex

\begin{document}
\baselineskip=18pt
\pagestyle{empty}
\begin{center}
\bigskip

\rightline{FERMILAB--Pub--96/222-A}
\rightline{August 1996}
\rightline{submitted to {\it Physical Review D}}

\vspace{.2in}
{\Large \bf Dynamical $\Lambda$ Models of Structure Formation}
\bigskip

\vspace{.2in}
Kimberly Coble$^{1,2}$\\
Scott Dodelson$^2$\\
Joshua A. Frieman$^{1,2}$\\
\vspace{.2in}
{\it ${}^1$Department of Astronomy and Astrophysics\\
The University of Chicago, Chicago, IL~~60637}\\
\vspace{.2in}
{\it ${}^2$NASA/Fermilab Astrophysics Center\\
Fermi National Accelerator Laboratory, Batavia, IL~~60510-0500}\\

\end{center}

\vspace{.3in}

\centerline{\bf ABSTRACT}
\bigskip
Models of structure formation with a cosmological constant
$\Lambda$ provide a good fit to the observed power spectrum of
galaxy clustering. However, they suffer from several problems. Theoretically,
it is difficult to understand why the cosmological
constant is so small in Planck units. Observationally, while the power
spectra of cold dark matter plus $\Lambda$ models
have approximately the right shape,
the COBE-normalized amplitude for a scale invariant
spectrum is too high, requiring galaxies to
be anti-biased relative to the mass distribution.
Attempts to address the first
problem have led to models in which a dynamical field supplies
the vacuum energy, which is thereby determined by
fundamental physics scales. We explore the implications of
such dynamical $\Lambda$ models for the formation of large-scale structure.
We find that there are dynamical models for which the
amplitude of the COBE-normalized spectrum matches the observations.
We also calculate the cosmic microwave background anisotropies in these
models and show that the angular power spectra are distinguishable from
those of standard cosmological constant models.

\newpage
\pagestyle{plain}
\setcounter{page}{1}
\newpage

\section{Introduction}

The cosmological constant has had a long and tortured history
since Einstein first introduced it in 1917 in order to obtain static
cosmological solutions \cite{jeremy}. Under observational duress,
it has been periodically invoked by cosmologists
and then quickly forgotten when the
particular crisis passed. Historical examples include the first `age crisis'
arising from Hubble's large value for the expansion rate (1929),
the apparent clustering of QSO's at a specific redshift (1967),
and early cosmological tests which indicated a negative deceleration
parameter (1974).

Recently, a cosmological model with substantial vacuum energy---a
relic cosmological constant $\Lambda$---has again come into vogue for
several reasons\cite{ostriker}.
First, dynamical estimates of the mass density on the scales of
galaxy clusters, the largest gravitationally bound systems, suggest
that $\Omega_m  = 0.2 \pm 0.1$ for the matter ($m$)
which clusters gravitationally (where $\Omega$
is the present ratio of the mean mass density of the universe to the critical
Einstein-de Sitter density, $\Omega = 8\pi G \rho/3H^2$) \cite{carlberg}.
However, if a sufficiently long epoch of inflation took place during
the early universe, the present spatial curvature should be negligibly
small, $\Omega_{tot} = 1$. A cosmological constant,
with effective density parameter $\Omega_\Lambda \equiv
\Lambda/3H^2_0 = 1 - \Omega_m$, is
one way to resolve the discrepancy between
$\Omega_m$ and $\Omega_{tot}$.

The second motivation for the revival of the cosmological constant is
the `age crisis' for spatially flat $\Omega_m = 1$
models. Current estimates of the Hubble
expansion parameter from a variety of methods appear to be
converging to
$H_0 \simeq 70 \pm 10$ km/sec/Mpc, while estimates
of the age of the universe from globular clusters are holding at
$t_{gc} \simeq 13 - 15$ Gyr or more. Thus, observations
imply a value for the the `expansion age'
$H_0 t_0 = (H_0/ 70 ~{\rm km/sec/Mpc})(t_0/14 ~{\rm Gyr}) \simeq 1.0 \pm 0.2$.
This is higher than that for the standard Einstein-de Sitter model
with $\Omega_m = 1$, for which $H_0 t_0 = 2/3$.
On the other hand, for models with a cosmological constant,
$H_0t_0$ can be larger. For example, for
$\Omega_\Lambda = 0.6 = 1-\Omega_m$, $H_0 t_0 = 0.89$.

Third, cold dark matter (CDM) models for
large-scale structure formation
which include a cosmological constant (hereafter, $\Lambda$CDM) provide
a better fit to the shape of the
observed power spectrum of galaxy clustering than
does the `standard' $\Omega_m=1$ CDM model \cite{lcdm}.
Figure 1 shows the inferred galaxy power spectrum today
(based on a recent compilation \cite{peacock}),
compared with the matter power spectra predicted by
standard CDM and a
$\Lambda$CDM model with $\Omega_\Lambda = 0.6 $. In both cases,
the Hubble parameter has been fixed to  $h \equiv
H_0/(100 $ km/sec/Mpc) $ =0.7$ and the baryon density to
$\Omega_B = 0.0255$, in the center of the range allowed by primordial
nucleosynthesis. Linear perturbation
theory has been used to calculate the model power spectra, $P(k)$,
defined by $\langle \delta({\bf k})\delta^*({\bf k}')\rangle
= (2\pi)^3 P(k) \delta_D({\bf k}-{\bf k}')$, where $\delta({\bf k})$
is the Fourier
transform of the spatial matter density fluctuation field and
$\delta_{D}$ is the Dirac delta function. Here and throughout,
we have taken the primordial power spectrum to be exactly scale-invariant,
$P_{primordial}(k) \propto k^{n}$ with $n=1$.
Standard CDM clearly gives a poor fit
to the shape of the observed spectrum \cite{standard},
while the $\Lambda$CDM model gives a good fit to the shape of the
observed spectrum. The amplitudes of the model
spectra in Fig. 1 have been fixed at large scales by observations of
cosmic microwave background (CMB) anisotropies by the
COBE satellite \cite{cobe,gorski}.

Despite these successes, cosmological constant models face several
difficulties of their own.
On aesthetic grounds, it is difficult to understand why
the vacuum energy density of the universe,
$\rho_\Lambda \equiv \Lambda/8\pi G$,
should be of order $(10^{-3} {\rm eV})^4$,
as it must be to have a cosmological impact ($\Omega_\Lambda \sim 1$).
On dimensional grounds, one would expect it to be many orders of magnitude
larger -- of order $m_{\rm Planck}^4$
or perhaps $m_{SUSY}^4$.  Since this is not the case,
we might plausibly assume that some physical mechanism sets the ultimate
vacuum energy to zero. Why then is it not zero today?

The cosmological constant is also increasingly observationally
challenged. Preliminary results from on-going searches\cite{perlmutter}
for distant Type Ia supernovae indicate that $\Omega_\Lambda <  
0.47$ (at 95\% confidence) 
for spatially flat $\Lambda$ models.
Furthermore, in $\Lambda$ models
a larger fraction of distant QSOs would be gravitationally
lensed than in a $\Lambda=0$ universe; surveys for lensed QSOs have
been used to infer the bound $\Omega_\Lambda \la 0.7$
\cite{kochanek}.

In this paper, we focus on a third problem of
cosmological constant models---the amplitude of the
power spectrum of galaxy clustering.
The shape of the $\Lambda$CDM power
spectrum in Figure 1 matches the galaxy power spectrum;
however the amplitude is too high. Indeed a number of
analyses have found that this problem persists on all
scales:
\begin{enumerate}

\item[$\bullet$] On the largest scales ($k < 0.1 {\rm h\ Mpc}^{-1}$),
linear theory should be adequate, and Figure 1 suggests that the amplitude
is too high by at least a factor of two.

\item[$\bullet$] On intermediate
scales, we can quantify the amplitude through the
dispersion of the density field smoothed over top-hat spheres of radius $R=
8 h^{-1}$ Mpc, denoted $\sigma_8$, where $\sigma^2(R) = 4\pi \int_0^\infty
k^2 P(k) W^2(kR) dk$, and $W(kR)$ is the Fourier-transform of the spatial
top-hat window function of radius $R$.
In the $\Lambda$CDM model of Figure 1, COBE normalization
yields $\sigma_8 \simeq 1.3$ \cite{gorski}, while galaxy surveys generally
indicate $\sigma_{8,gal} \simeq 1$ for optically selected galaxies
and $\sim 0.8$ for galaxies selected by infrared flux. This high
COBE normalization also marginally conflicts with the abundance of rich galaxy
clusters \cite{wef}. Using the observed cluster X-ray temperature
distribution function and modelling cluster formation using
Press-Schechter theory, for this $\Lambda$CDM model the cluster
abundance implies $\sigma_8 \simeq 1.0 ^{+.35}_{-.26}$ \cite{vl},
where the errors are approximate 95\% confidence limits.

\item[$\bullet$]  N-body simulations indicate that
the power spectrum amplitude is higher by a factor of two
to three than that found in galaxy surveys at small scales, $k \ga
0.4 h$ Mpc$^{-1}$ \cite{primack}.
Thus, the cosmological constant model
would require galaxies to be substantially {\it anti-biased}
with respect to the mass distribution, $\sigma_{gal} < \sigma_\rho$.
Models of galaxy formation, however, suggest that the bias
parameter, $b \equiv \sigma_{gal}/\sigma_\rho$, is greater than unity
\cite{mowhite,tilt}.

\end{enumerate}

Motivated by these difficulties,
we consider models in which the energy density
resides in a dynamical scalar field rather than in a pure vacuum
state. These {\it dynamical}
$\Lambda$ models \cite{fhsw,dynlam}\
were proposed in response to the aesthetic difficulties of cosmological
constant models. They were also found \cite{fhsw} to partially
alleviate their observational problems as well; for example, the 
statistics of gravitationally lensed QSOs yields a less restrictive 
upper bound on $H_0 t_0$ in these models\cite{dynlens}.  
We emphasize here that they may also solve
the galaxy clustering amplitude problem.

To get a preview of this conclusion,
Fig. 1 also shows the COBE
normalized power spectrum for a dynamical $\Lambda$ model with present
scalar field density parameter $\Omega_\phi = 0.6$ (see
\S 3 for a discussion of these models).
While the shape of the spectrum is identical to that of
the $\Lambda$CDM model with $\Omega_\Lambda = 0.6$,
the scalar field model has a lower amplitude,
and thus provides a better fit to the galaxy clustering data.
In \S 2, we explain these features of the power spectrum for the standard
$\Lambda$CDM model and for generic dynamical $\Lambda$ models.
The remaining sections investigate in detail
a specific class of models as a worked example.
Section 3 reviews the scalar field model, based on ultra-light
pseudo-Nambu-Goldstone bosons (PNGBs) \cite{fhsw}.
To explore the parameter space of this model, we have adapted a code which
solves the linearized Einstein-Boltzmann equations for perturbations
to a Friedmann-Robertson-Walker (FRW) background. The appendices
contain details of these modifications. Section 4 discusses the
qualitative features of cosmic evolution in the PNGB models and
presents results of our calculation for the amplitude of the power
spectrum in this model. In \S 5 we present the cosmic microwave
background (CMB) power spectrum for a particular set of model parameters,
followed by the conclusion.

\section{The Power Spectrum}

\subsection{The Shape of $P(k)$}

Figure 1 suggests that standard
CDM could be improved by simply shifting the
turnover in the power spectrum to larger scales (smaller wavenumber $k$).
This is a plausible fix, for the location of the turnover
corresponds to the scale that entered the
Hubble radius when the universe became matter-dominated. On scales
smaller than this, the fluctuation amplitude is
suppressed compared to that on larger scales, because matter perturbations
inside the Hubble radius
cannot grow in a radiation-dominated universe. This scale is
determined by the ratio of matter to radiation
energy density at early times.
To ``fix'' CDM, one must decrease the ratio $\bar\rho_m/\bar\rho_r$
in the universe today below that predicted by the
standard Einstein-de Sitter model. The matter and radiation densities
scale as $\bar\rho_m=\bar\rho_{m,0} a^{-3}$
and $\bar\rho_r=\bar\rho_{r,0} a^{-4}$,
where the cosmic scale factor $a$ is normalized to unity today
($a_0=1$)
and the subscript $0$ denotes the present. Thus
the epoch of matter-radiation equality
is determined by the present energy densities of matter and radiation:
\begin{equation}
a_{EQ} = {\bar\rho_{r,0}\over \bar\rho_{m,0}}
= {4.3\times 10^{-5} \over\Omega_m h^2} ~~.
\label{aeq}
\end{equation}
Decreasing the
matter to radiation density ratio shifts the epoch of matter-radiation equality
closer to the present, thereby moving the turnover in the power spectrum
to larger scales.

Indeed, this shift
is precisely what is done in several currently popular models
of structure formation. Examples include
i) models with a lower Hubble constant than indicated by
observations\cite{silk}, ii) models with extra relativistic
degrees of freedom\cite{taucdm}, and
iii) models with a cosmological constant\cite{lcdm}. Since
$\bar\rho_m \propto \Omega_m h^2$, a lower Hubble constant
decreases the ratio of
matter to radiation density today. Adding more relativistic degrees of
freedom adds to the radiation content, decreasing
the ratio of matter to radiation. Finally, in spatially flat
$\Lambda$ models, $\Omega_m \equiv 1 - \Omega_{\Lambda}$
is reduced from its standard CDM value ($\Omega_m = 1$), achieving
a similar effect.

Thus the main benefit of $\Lambda$ models
for the shape of the power spectrum is
that $\Omega_{m}$ is smaller than in the standard CDM model.
For the purpose of
the power spectrum shape, the value of the
vacuum energy density at early times is irrelevant, as
long as it is negligible compared to the matter and radiation densities
at matter-radiation equality. While the time dependence of the
vacuum energy density is different for various
dynamical $\Lambda$ models, all such models yield
the same power spectrum shape
for a fixed value of the present vacuum energy density.
We emphasize this point in Fig. 2,
which shows the energy densities of matter, radiation, $\Lambda$,
and a specific dynamical $\Lambda$ model (scalar field $\phi$),
as a function of scale factor $a$. With $\Omega_\Lambda$ and $\Omega_\phi
=0.6$ today, the standard and dynamical $\Lambda$ models have the
same shape for $P(k)$ (shown in Fig. 1),
since they have identical values of $a_{EQ}$.
As we will see below, however, the amplitudes of the power spectra
in these models differ substantially.

\subsection{The Amplitude of $P(k)$}

Compared to standard CDM,
three new physical effects \cite{whitebunn}\
conspire to change the amplitude of the matter
power spectrum in COBE-normalized
$\Lambda$ models: i) the suppression of growth
of perturbations when the universe becomes $\Lambda$-dominated,
ii) the reduced gravitational potential, and iii)
the integrated Sachs-Wolfe (ISW) effect. We review these effects
in turn.

The equations governing large scale perturbations in a
flat universe with matter and  vacuum energy
are
\begin{equation}
\ddot \delta +  Ha \dot \delta
- {3\over 2a^2} H^2 \Omega_m \delta = 0
\label{pert}
\end{equation}
\begin{equation}
 H^2
= {H_0^2\over a^3} \left[ \Omega_m + \Omega_\Lambda {\rho_\Lambda\over
\rho_{\Lambda,0}} a^3 \right]
\label{einstein}
\end{equation}
Here overdots denote derivatives with respect to conformal
time $\tau$, where $\tau \equiv \int dt/a(t)$, $\rho_\Lambda$
is the vacuum energy density, not necessarily equal to
its present value $\rho_{\Lambda,0}$, the density fluctuation amplitude
$\delta({\bf x},\tau) \equiv (\rho_m({\bf x},\tau) - \bar\rho_m(\tau))/
\bar\rho_m(\tau)$, and $H$
is the Hubble expansion rate [we use units in which $\hbar=c=1$].

Equation \ref{pert} essentially describes the behavior of a
damped harmonic oscillator.
When the energy density of the
universe becomes dominated by a $\Lambda$ or
dynamical $\Lambda$, i.e., the second term on the the RHS in equation
\ref{einstein}\ becomes important, the damping becomes more severe.
When this happens, the growth of perturbations is suppressed.
As a function of $\Omega_{m}$, this suppression
can be described by the scaling,
\begin{equation}
{\delta_{0}/\delta_{(z=100)}} \propto \Omega_{m}^p ,
\label{suppress}
\end{equation}
where $\delta_{0}$ is the perturbation amplitude today, and $\delta_{(z=100)}$
is the amplitude at the epoch $z \equiv (1/a)-1=100$, chosen as
an arbitrary early epoch before the vacuum energy becomes dynamically
important. In $\Lambda$CDM models, $p\simeq 0.2$. For dynamical
$\Lambda$ models, the suppression exponent depends on the details
of the specific model, but it is generally greater than that in
$\Lambda$CDM models, because the dynamical $\Lambda$ dominates
earlier in the history of the universe for fixed $\rho_{\Lambda,0}$.
For the model shown in Fig. 1, $p\simeq 0.56$. For 
open CDM models (with $\Lambda=0$), the scaling is also 
$p \simeq 0.56$.

As a result of this suppression, one might expect
the amplitude of the power spectrum in $\Lambda$CDM and dynamical
$\Lambda$ models to be smaller than that
in standard CDM. However, from the Poisson equation
\begin{equation}
 \nabla^2 \Phi =  {3\over 2a^2} H^2\Omega_m \delta ~,
\label{poisson}
\end{equation}
we have $\Phi \propto \Omega_{m} \delta$, where $\Phi$ is the
gravitational potential associated with large-scale density fluctuations.
Since the CMB anisotropy at large angle is a
well-defined function of the potential\cite{husugiyama},
COBE normalization corresponds to fixing the potential, i.e., to
fixing $\Omega_{m}\delta$.
For COBE-normalized models,  the growth suppression
and Poisson's equation combine to yield the scale-independent
relation $\delta \propto \Omega_{m}^{p-1}$. Thus
the power spectrum $P(k) \propto \delta^{2} \propto \Omega_{m}^{-1.6}$
in $\Lambda$CDM models. A larger cosmological constant
implies a smaller $\Omega_{m}$, which in turn implies a larger
amplitude for the power spectrum.
In dynamical $\Lambda$ models, $p$ is not fixed at $0.2$,
so the amplitude of the power spectrum can be smaller than in standard
$\Lambda$ models. For the model of Fig. 1, with $p=0.56$,
$P(k) \propto \Omega_{m}^{-0.9}$.

The integrated Sachs-Wolfe effect (ISW), which is due
to time evolution of the  potential, also affects the amplitude
of the power spectrum. The changing potential at late times
in $\Lambda$ models increases the anisotropy on the large
angular scales probed by COBE.
Thus, for fixed COBE normalization, the amplitude
of the power spectrum decreases, changing the
dependence of the power spectrum on $\Omega_{m}$ to
$P \propto \Omega_{m}^{-1.4}$ in the $\Lambda$CDM model. In dynamical
$\Lambda$ models, where the potential typically changes more
than in standard $\Lambda$ models, the ISW effect tends to be
larger and is not a power law function of $\Omega_{m}$. Hence the
power spectrum amplitude in dynamical models
is even less enhanced than in $\Lambda$CDM models, and
can even be reduced compared to standard CDM.

\section{Ultra-light Scalar Fields}

A number of models with a dynamical $\Lambda$
have been discussed in the literature
\cite{dynlam}. We will focus on a particular class of models motivated
by the physics of pseudo-Nambu-Goldstone bosons
(hereafter PNGBs) \cite{fhsw,hr}.

It is conventional to assume that the fundamental vacuum energy
of the universe is zero, owing to some as yet not understood
mechanism, and that this mechanism
`commutes' with other dynamical effects that lead
to sources  of energy density. This
is required so that, e.g., at earlier epochs
there can temporarily
exist non-zero vacuum energy which allows inflation to take place.
With these assumptions, the effective
vacuum energy at any epoch will be dominated by the heaviest fields
which have not yet relaxed to their vacuum state. At late times, these
fields must be very light.

Vacuum energy is most simply stored in the
potential energy $V(\phi) \sim M^4$ of a scalar field, where $M$ sets the
characteristic height of the potential, and we set $V(\phi_m)=0$ at the
minimum of the potential by the assumptions above.
In order to generate a non-zero
$\Lambda$ at the present epoch, $\phi$ must initially be displaced from the
minimum ($\phi_i \neq \phi_m$ as an initial condition), and its
kinetic energy must be small compared to its potential energy.
This implies that the motion of the field
is still overdamped, $m_\phi \equiv \sqrt{|V''(\phi_i)|}
\la 3H_0 = 5\times 10^{-33} h$ eV. In addition, for $\Omega_\Lambda \sim 1$,
the potential energy density should be of order the critical density,
$M^4 \sim 3H^2_0 M^2_{Pl}/8\pi$, or $M \simeq 3\times 10^{-3}h^{1/2}$ eV.
Thus, the characteristic height and curvature of the potential are
strongly constrained for a classical model of the cosmological constant.

This argument raises an apparent difficulty for such a model:
why is the mass scale $m_\phi$ thirty orders of magnitude smaller than $M$?
In quantum field theory, ultra-low-mass scalars are not {\it generically}
natural: radiative corrections generate large mass renormalizations at
each order of perturbation theory. To incorporate ultra-light scalars
into particle physics, their small masses should be at least
`technically' natural, that is, protected
by symmetries, such that when the small masses are set
to zero, they cannot be generated in any order of perturbation theory,
owing to the restrictive symmetry.

From the viewpoint of quantum field theory, PNGBs are the simplest way to have
naturally ultra--low mass, spin--$0$ particles.
PNGB models are characterized by two mass scales, a spontaneous symmetry
breaking scale $f$ (at which the effective Lagrangian
still retains the symmetry) and
an explicit breaking scale $\mu$ (at which the effective Lagrangian contains
the explicit symmetry breaking term). In terms of the mass scales
introduced above, generally $M \sim \mu$ and the PNGB mass
$m_\phi \sim \mu^2/f$. Thus, the two
dynamical conditions on $m_\phi$ and $M$ above
essentially fix these two mass scales to be $\mu \sim M \sim 10^{-3}$ eV,
interestingly close to the neutrino mass scale for the MSW solution to
the solar neutrino problem, and $f \sim M_{Pl} \simeq 10^{19}$ GeV,
the Planck scale. Since
these scales have a plausible origin in particle
physics models, we may have an explanation for the `coincidence' that the
vacuum energy is dynamically important at the present epoch.
Moreover, the small mass $m_\phi$ is technically natural.

An example of this phenomenon is the `schizon' model \cite{hr}, based on
a $Z_N$-invariant low-energy effective chiral
Lagrangian for $N$ fermions, e.g., neutrinos,
with mass of order $M$, in which the small PNGB mass,
$m_{\phi} \simeq M^{2}/f$,
is protected by fermionic chiral symmetries.
The potential for the light scalar field $\phi$ is of the form
\begin{equation}
V(\phi) = M^4 [\cos(\phi/f)+1] ~~.
\end{equation}
Since $\phi$ is
extremely light, we assume that it is the only classical field which
has not yet reached its vacuum expectation value.
The constant term in the PNGB potential has been chosen to ensure
that the vacuum energy vanishes at the minimum of the $\phi$
potential, in accord with our assumption that the fundamental 
vacuum energy is zero. 

\section{Cosmic Evolution and Large-scale Power Spectrum in PNGB Models}

To study the cosmic evolution of these models, we focus on the
spatially homogeneous, zero-momentum mode of the field,
$\phi^{(0)}(\tau) = \langle \phi({\bf x}, \tau) \rangle$, where the
brackets denote spatial averaging. We are
assuming that the spatial fluctuation amplitude $\delta \phi({\bf x},\tau)$
is small compared
to $\phi^{(0)}$, as would be expected after inflation if the post-inflation
reheat temperature $T_{RH} < f \sim M_{pl}$.
The scalar equation of motion is given in Appendix A.

The cosmic evolution of $\phi$ is determined by the ratio of its
mass, $m_\phi \sim M^2/f$, to the instantaneous expansion rate, $H(\tau)$.
For $m_\phi \la 3H$, the field evolution is overdamped by the expansion, and
the field is effectively frozen to its initial value $\phi_i$. Since $\phi$ is
initially laid down in the early universe (at a temperature $T \sim f \gg
M$) when its potential was dynamically
irrelevant, its initial value in a given Hubble volume
will generally be displaced from its vacuum
expectation value $\phi_m = \pi f$ (vacuum misalignment). Thus, at
early times, the field acts as an effective cosmological constant, with
vacuum energy density and pressure
$\rho_{\phi} \simeq - p_\phi \sim M^4$. At late times,
$m_\phi \gg 3H(\tau)$, the field undergoes damped oscillations about the
potential
minimum; at sufficiently late times, these oscillations are approximately
harmonic, and the stress-energy tensor of $\phi$ averaged over an oscillation
period is that of non-relativistic matter, with energy density
$\rho_\phi \sim a^{-3}$ and pressure $p_\phi \simeq 0$.

Let $\tau_x$ denote the epoch
when the field becomes dynamical, $m_\phi = 3H(\tau_x)$, with corresponding
redshift $1+z_x = 1/a(\tau_x))=(M^2/3H_0f)^{2/3}$. For comparison,
the universe makes the transition from radiation- to
matter-domination at $z_{eq} \simeq 2.3\times 10^4 \Omega_m h^2$,
much earlier than when the field becomes dynamical.
The $f-M$ parameter space is shown
in Fig. 3. In the far right portion of the figure, the field
becomes dynamical before the present epoch and currently redshifts as
non-relativistic matter; on the far left, $\phi$ is still frozen
and acts as an ordinary cosmological constant.
In the dynamical region, the present
density parameter for the scalar field is approximately
$\Omega_\phi \sim 24 \pi(f/M_{Pl})^2$, independent of
$M$ \cite{fhw}. The quasi-horizontal lines show contours of
constant $\Omega_\phi$, assuming a typical initial field value
$\phi_i/f = 1.6$ (we will use this value of $\phi_i/f$
for all the plots below; the quoted limits and results depend slightly
on it). The limit $\Omega_\phi < 1$ corresponds
approximately to $f< 3.5\times 10^{18}$ GeV.
In the frozen region, on the other
hand, $\Omega_\phi$ is determined by $M^4$, independent of $f$, and the
contours of constant $\Omega_\phi$ are nearly vertical. In this region, the
bound $\Omega_\phi < 1$ corresponds roughly to $M <0.003$ eV.

Figure 4 shows contours of constant $H_0 t_0$ in the same parameter space.
As expected, models with large $H_0t_0$ are concentrated toward the left hand
portion of the figure; as one moves to the right, $H_0 t_0$ asymptotically
approaches the Einstein-de Sitter value $2/3$, since the scalar field currently
redshifts as non-relativistic matter and we have assumed a spatially
flat universe. Consequently, the `interesting'
region of parameter space is the area near
the `corner' in Figs. 3 and 4, in which the field
becomes dynamical at recent epochs, $z_x \sim 0-3$. This
has new consequences, compared to $\Lambda$ models, for the
classical cosmological tests,
the expansion age $H_0t_0$, and large-scale structure.
In this region, the mass of the PNGB field is miniscule,
$m_\phi \sim 3H_0 \sim 4\times 10^{-33}$ eV, and (by construction) its
Compton wavelength is  of order the current Hubble radius, $\lambda_\phi =
m^{-1}_\phi =H_0^{-1}/3 \sim 1000 h^{-1}$ Mpc.

Figure 5 shows contours of the amplitude of galaxy clustering in the
$f-M$ parameter space. The amplitude shown is the quantity
\begin{equation}
\lim_{k \rightarrow 0} \left[{(P(k)/k)_\phi \over (P(k)/k)_\Lambda}\right] ,
\end{equation}
i.e., the amplitude on large scales relative to that
for a $\Lambda$CDM model with
the same effective density as the PNGB model, $\Omega_\Lambda = \Omega_\phi$.
This amplitude ratio goes to unity in the left-hand portion of the
figure since that region corresponds to a 
$\Lambda$CDM model. However the amplitude ratio can
be substantially below one in the dynamical region on the right. The
cross marks the specific choice $M=0.005$ eV,
$f=1.885 \times 10^{18}$ GeV, with initial
field value $\phi_i/f=1.6$, yielding
$\Omega_{\phi}=0.6$, which corresponds to the parameters used for the
dynamical $\Lambda$ curves in
Figs. 1 and 2. For this case, the X-ray cluster abundance yields
$\sigma_8^{cl} \simeq 0.9^{+.3}_{-.2}$, in good agreement with
the COBE normalization $\sigma_8^{COBE} \simeq 0.8$ for this model.
Figure 6 shows how density perturbations grow in the different models. 
From Eqn.(4) and the text following, 
the dynamical $\Lambda$ model has a higher amplitude at early times
than a $\Lambda$CDM model with the same amplitude today. As a 
consequence, there
should be no problem accounting for high-redshift objects such as 
QSOs and Lyman-alpha clouds in this
model. 

Note that the factor $\delta(z)/\delta_0$, relative to its 
value in the standard CDM model, approaches $\Omega_m^{-p}$ at 
$z \gg 1$, where $p$ is the scaling exponent discussed in \S 2. 
As a result, the non-linear behavior of the 
dynamical $\Lambda$ model follows that of
an open model with the same value of $\Omega_m$. 
We estimate
the non-linear behavior by using the fitting formula of
Ref. \cite{pdf}, following the original treatment \cite{hamilton}\
of Hamilton et al.
Figure 7 shows these non-linear spectra. 
On scales $k \leq 1 h {\rm Mpc}^{-1}$,
the amplitude of the power spectrum is indeed a
factor of two smaller in the dynamical
$\Lambda$ model than in the corresponding $\Lambda$CDM model.

We note by comparing Figs. 4 and 5 that the region of parameter
space in which the amplitude (anti-bias) problem is solved, i.e., in which
the amplitude ratio is approximately in the range $0.3 - 0.5$, is the
one in which the age of the universe is only slightly greater than
in the Einstein-de Sitter $\Omega_m=1$ case. For our specific
model above, $H_{0}t_{0}=0.73$. For the corresponding
$\Lambda$CDM model with the same value of $\Omega_m$, 
$H_{0}t_{0}=0.89$, more comfortably
within the observational limits. This is a general feature of 
the dynamical models considered here: for fixed $\Omega_m$, the 
standard $\Lambda$ model gives an upper bound on $H_0t_0$. 
Thus, the amplitude problem in this model is resolved partially 
at the expense of the age problem. On the other hand, the $q_0$ 
constraints from SNe and gravitational lensing translate into 
weaker upper bounds on $H_0t_0$ for the dynamical as opposed to 
the standard $\Lambda$ models.  
Although we have not thorougly examined all models, it is clear that 
one could explore the PNGB model parameter space to obtain a more balanced 
compromise between the age problem and the anti-bias problem. 
For example, for $f\simeq 2.5\times 10^{18}$ GeV and $M \simeq 0.0035$ eV, 
the amplitude ratio is about 0.5, and one has $\Omega_\phi \simeq 0.75$ and
$H_0 t_0 \simeq 0.9$. In this case, with $h=0.7$, the power spectrum 
shape is reasonable ($\Omega_m h \simeq 0.15$) and the age of the universe 
is $t \simeq 12.6$ Gyr. 

Comparing Figs. 3 and 5, and focusing on the dynamical region near the
`corner' of the parameter space, we see that the power spectrum shape and
amplitude constraints fix the free parameters of the model.
That is, as noted in \S 2, the shape of the spectrum is
fixed by requiring $\Omega_\phi \simeq 0.6$, which determines the
scale $f$. Near the corner, fixing the amplitude then determines the
other mass scale $M$. While these figures correspond to a specific
choice of the initial field value $\phi_i/f$, the scalar field evolution
is universal in the sense that
a shift in the mass scale $f$, accompanied by an appropriate rescaling of
$\phi_i$, leads to essentially identical evolution. Consequently,
compared to $\Lambda$CDM models, these dynamical models
have only one additional free parameter, the mass $M$, to solve
the amplitude (anti-bias) problem.

\section{CMB Anisotropy}

The angular power spectra of the cosmic microwave
background (CMB) anisotropy for dynamical $\Lambda$ models
are distinguishable from those of standard CDM and $\Lambda$CDM
models. CMB angular power is usually expressed in terms
of the angular multipoles $C_{l}$. If the sky temperature is expanded
in terms of spherical harmonics as
$T(\theta,\phi)=\Sigma_{lm}a_{lm}Y_{lm}(\theta,\phi)$, then
$C_{l}=\langle |a_{lm}|^{2}\rangle$, where large $l$
corresponds to small angular scales.
The angular power spectra for standard CDM ($\Omega_m=1$), $\Lambda$CDM,
and dynamical $\Lambda$ models (the latter two with $\Omega_m=0.4$)
are shown in Figures 8 and 9 for $h=0.7$,
$\Omega_B = 0.0255$, and primordial spectral index
$n=1$. Following standard practice, we plot the
product $l(l+1)C_l$, normalized to its value at $l=10$,
vs. $l$.

The Appendices contain the details of the alterations required in the standard
Boltzmann code to calculate the CMB anisotropy in scalar field dynamical
$\Lambda$ models. We can, however, identify two physical
effects primarily responsible for the differences in the CMB
signature between the $\Lambda$CDM and dynamical $\Lambda$ models shown in
Figs. 8 and 9. First, the present ages in
conformal time coordinates, $\tau_0$, are different in the two models.
Even though the acoustic oscillations responsible
for the peaks in the CMB angular spectrum
occur at the same physical scales (or same Fourier
wave numbers $k$), the correspondence between $k$ and angular multipole
$l$ differs. Typically, in a flat universe, a given multipole $l$\
corresponds to a fixed value of $k\tau_0$. Thus,
the dynamical $\Lambda$ angular
spectra are shifted in $l$ by the ratio of the present conformal times
in the two models.
Second, since the scalar field evolves at late times,
the gravitational potential changes more rapidly in the dynamical
$\Lambda$ model. This leads to an enhanced ISW effect, and
therefore a relatively larger $C_l$ at large scales (low $l$),
as shown in Fig. 9.
Thus, for models normalized by COBE, which approximately fixes
the spectrum at $l \simeq 10$, the angular amplitude $l(l+1)C_l$
at small scales (large $l$) is smaller in the dynamical $\Lambda$ model.

\section{Conclusions}

The observational arguments in favor of the resurrection of
the cosmological constant apply to dynamical $\Lambda$ models as well.
In addition, the dynamical $\Lambda$ models offer a potential physical
explanation for the curious coincidence that $\Omega_\Lambda$
is close to one, by relating the present vacuum energy density to
mass scales in particle physics.  In the ultra-light 
pseudo-Nambu-Goldstone boson models, this is 
achieved through  spontaneous symmetry breaking near the Planck 
scale, $f \sim M_{Pl}$, and explicit breaking at a scale 
reminiscent of MSW neutrino masses, $M \sim 10^{-3}$ eV. In combination 
with the assumption that the true vacuum energy vanishes (due to 
an as yet unknown physical mechanism), such a model provides an 
example of a dynamical $\Lambda$. 

We have shown
that such dynamical models can lead to a lower amplitude
for density fluctuations compared to standard $\Lambda$ models, thereby
alleviating the anti-bias problem. The advantages of the cosmological 
constant for the shape of the power spectrum are retained in the 
dynamical models as well.
Such dynamical models are, moreover,
distinguishable from constant-$\Lambda$ models by virtue
of their CMB angular spectra.

\section*{Acknowledgments}
We thank Andrew Liddle and Martin White for useful conversations.
This work was supported in part by the DOE (at
Fermilab) and the NASA (at Fermilab through grant NAG 5-2788).

\begin{appendix}

\section{Changes in standard Boltzmann code}

This Appendix and the following briefly outline the new physics
incorporated into the Boltzmann code in $\Lambda$CDM and
dynamical $\Lambda$ models.
Since the Hubble parameter is determined by the sum over
densities of all species, $H^2=(8\pi G)\Sigma_{i}\rho_{i}$,
inclusion of a cosmological constant $\Lambda$ or scalar field
$\phi$ changes the relationship between the cosmic scale factor $a$
and conformal time $\tau$, since $(da/d\tau)/a^2=H$.
In addition to the species
included in the standard Boltzmann code, namely,
baryons, cold dark matter, photons, and three massless neutrinos,
the density in a cosmological constant or scalar field $\phi$ is
now included. In $\Lambda$CDM models, the vacuum energy
density $\rho_\Lambda = \Lambda/8\pi G$ is constant. In the
dynamical models, the scalar field energy density $\rho_\phi$
can be solved for with the scalar equation of motion for the
homogeneous part $\phi^{(0)}(\tau)$ of the field,

\begin{equation}
\ddot \phi^{(0)} + 2Ha \dot \phi^{(0)} + a^{2} dV(\phi^{(0)})/d\phi^{(0)}
= 0 ~~,
\label{zero-order-EOM}
\end{equation}
where the scalar field potential is
\begin{equation}
V(\phi)=M^{4}[\cos(\phi/f)+1] ~~,
\label{phi-potential}
\end{equation}
and the scalar energy density
\begin{equation}
\rho_{\phi}={1 \over 2a^2} \dot \phi^{(0)2} + V(\phi^{(0)}) ~~.
\label{zero-order-phi}
\end{equation}
Here overdots denote derivatives with respect to conformal time $\tau$.

\section{Perturbation equations for dynamical $\Lambda$ models}

The general equation of motion for the scalar field
$\phi({\bf x},\tau)$ is derived by minimizing the action

\begin{equation}
S = \int d^4x \sqrt{-g} \left[
        {1 \over 2} g_{\mu\nu} \partial^\mu \phi \partial^\nu \phi
        - V(\phi)  \right]
\label{action}
\end{equation}
with respect to variations in $\phi$.
The metric is that of a perturbed Friedmann-Robertson-Walker universe,
\begin{equation}
g_{\mu\nu}({\bf x}, \tau) = g^{(0)}_{\mu\nu}(\tau) + \delta g_{\mu\nu}({\bf
x}, \tau) ~~,
\label{metric}
\end{equation}
where $g^{(0)}_{\mu\nu}$ is the homogeneous part which describes the
Hubble expansion, and $\delta g_{\mu\nu}$ is the metric perturbation.
In synchronous gauge, the latter can be
parametrized by the variables $h,h_{33}$ as in \cite{efstathiou}.
The scalar field can be similarly decomposed into a homogeneous
part and a spatial perturbation,

\begin{equation}
\phi({\bf x},\tau) = \phi^{(0)}(\tau) + \delta\phi({\bf x}, \tau) ~~,
\label{phipert}
\end{equation}
where $\phi^{(0)}$ is the solution to the spatially homogeneous
equation of Appendix A. Keeping only terms linear in
$h,h_{33}$, and $\delta\phi$, and taking the Fourier transform
yields the equation of motion for the Fourier amplitude $\delta\phi_k$,

\begin{equation}
\ddot {(\delta \phi_k)} + 2Ha \dot {(\delta \phi_k)}
  + \left(k^{2} + a^{2}[d^{2}V/d\phi^{2}]_{\phi=\phi^{(0)}(\tau)}\right)
{(\delta \phi_k)}  = {\dot h \dot \phi^{(0)} \over 2}
\label{first-order-phi}
\end{equation}

There will also be an additional source term in the
Einstein equation for the metric perturbation. Again following the notation
of \cite{efstathiou}, the Einstein equation becomes:

\begin{equation}
\ddot h + Ha \dot h = 8 \pi G \left(S_\phi + S_u \right)
\label{einstein-h}
\end{equation}
where the source term due to $\phi$ is given by
\begin{equation}
S_\phi = 4 \dot {(\delta \phi)} \dot \phi
   -2 a^{2}{(\delta \phi)} [dV/d\phi]_{\phi=\phi^{(0)}(\tau)} ~~,
\label{phisource}
\end{equation}
and $S_u$ contains the usual source terms for matter and
radiation \cite{efstathiou}.

\end{appendix}


\clearpage

\begin{figure}[t!]

\centering

\centerline{\epsfxsize=15. truecm \epsfysize=10. truecm 
\epsfbox{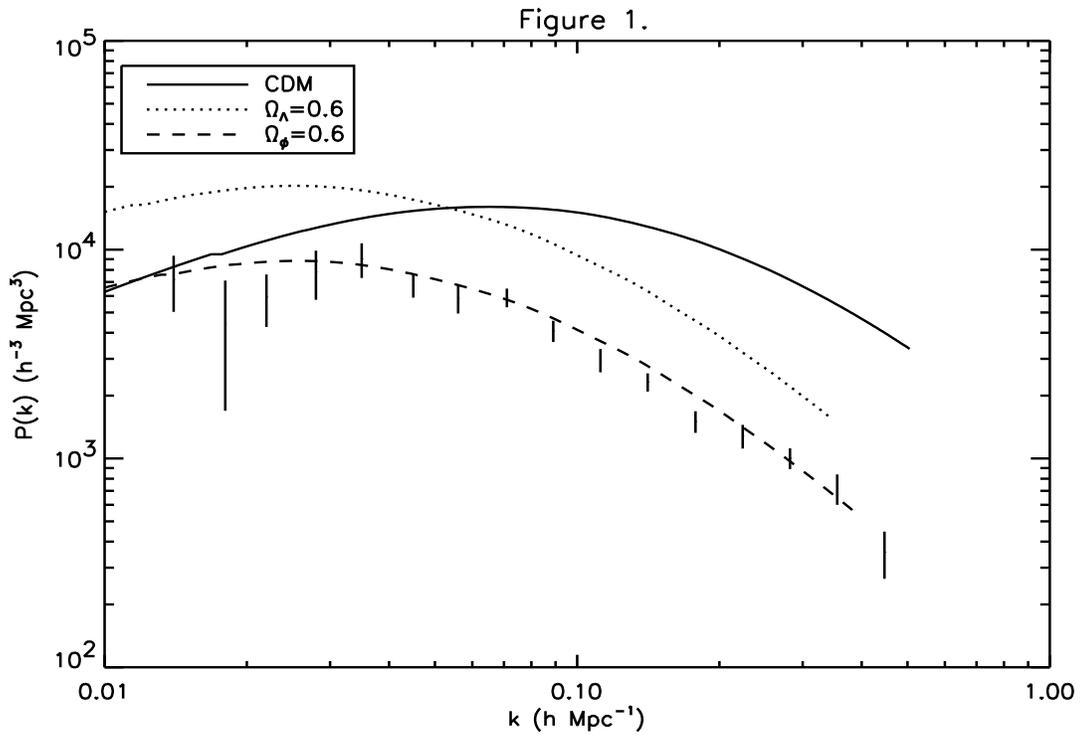}
}

\vspace{1. truecm}


\caption{COBE-normalized power spectra for standard 
CDM, $\Lambda$CDM with $\Omega_{\Lambda}=0.6$,
and scalar field $\Omega_{\phi}=0.6$ models.
In all models $h=0.7$, $\Omega_{B}=0.0255$, and $n=1$.
The data points are based on a recent compilation of
galaxy clustering data by Peacock and Dodds \cite{peacock}.}
\label{fig1}
\end{figure}

\vspace{1. truecm}

\begin{figure}[t!]

\centering

\centerline{\epsfxsize=15. truecm \epsfysize=10. truecm 
\epsfbox{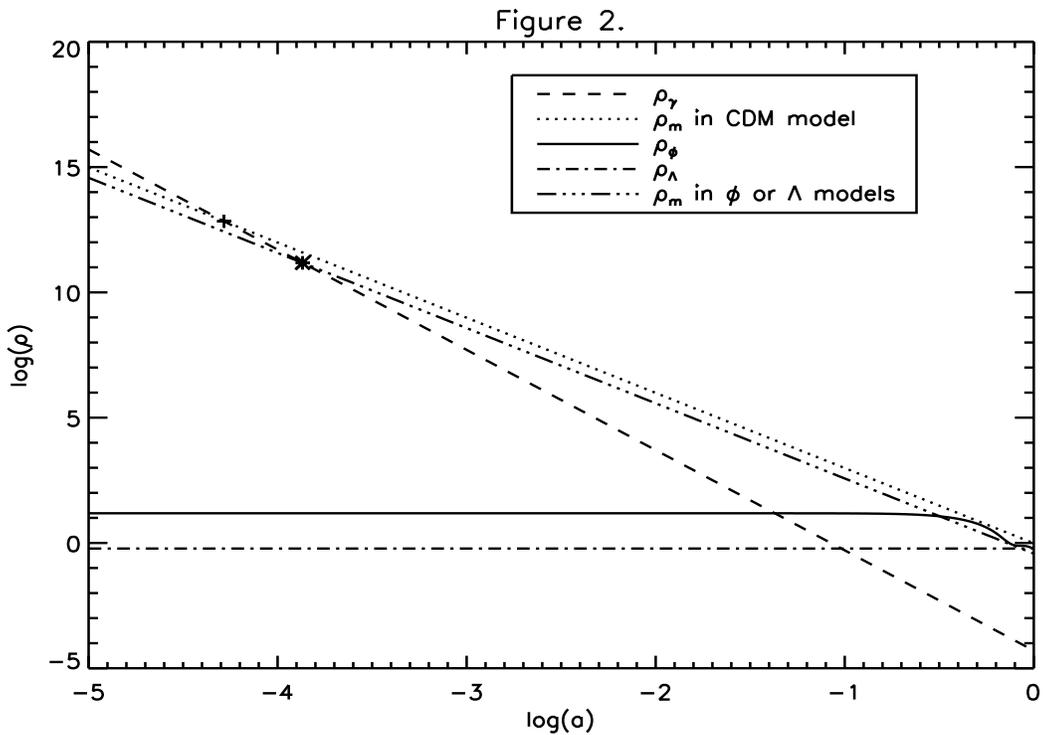}
}

\vspace{1. truecm}


\caption{Density ${\bar \rho}$ vs. cosmic scale factor $a$.
Fixing $\Omega_{\Lambda}$ or $\Omega_{\phi}$
to 0.6 lowers $\Omega_{m}$ from the standard CDM
value of 1.0, pushing the epoch of
matter-radiation equality, $a_{EQ}$, closer to today.
The cross denotes $a_{EQ}$ for the standard CDM model and
the asterisk denotes $a_{EQ}$ for the $\Lambda$ and $\phi$
models.}
\label{fig2}
\end{figure}

\clearpage

\begin{figure}[t!]

\centering

\centerline{\epsfxsize=15. truecm \epsfysize=10. truecm 
\epsfbox{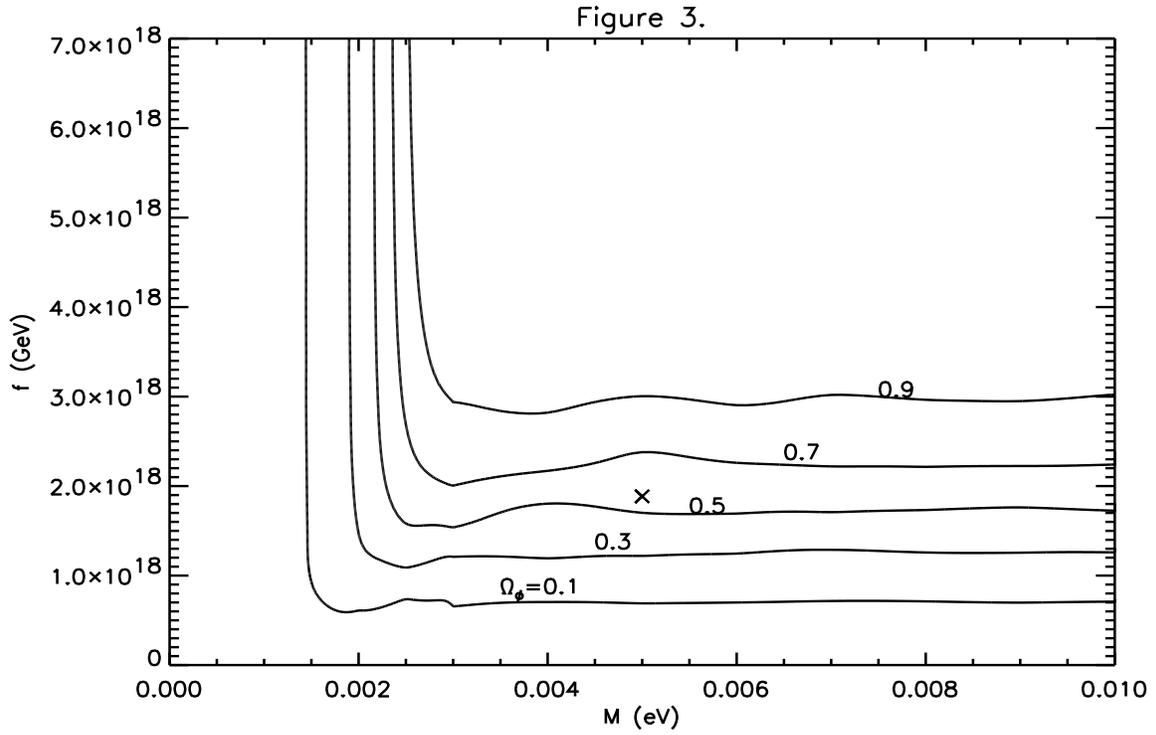}
}

\vspace{1. truecm}


\caption{Contours of $\Omega_{\phi}$ in the PNGB parameter
space, assuming an initial field value $\phi_i/f = 1.6$.
The cross marks the choice $M=0.005$ eV,
$f=1.885 \times 10^{18}$ GeV, yielding
$\Omega_{\phi}=0.6$, which is the model shown in Figures
1, 2, and 6 -- 9.}
\label{fig3}
\end{figure}

\clearpage

\begin{figure}[t!]

\centering

\centerline{\epsfxsize=15. truecm \epsfysize=10. truecm 
\epsfbox{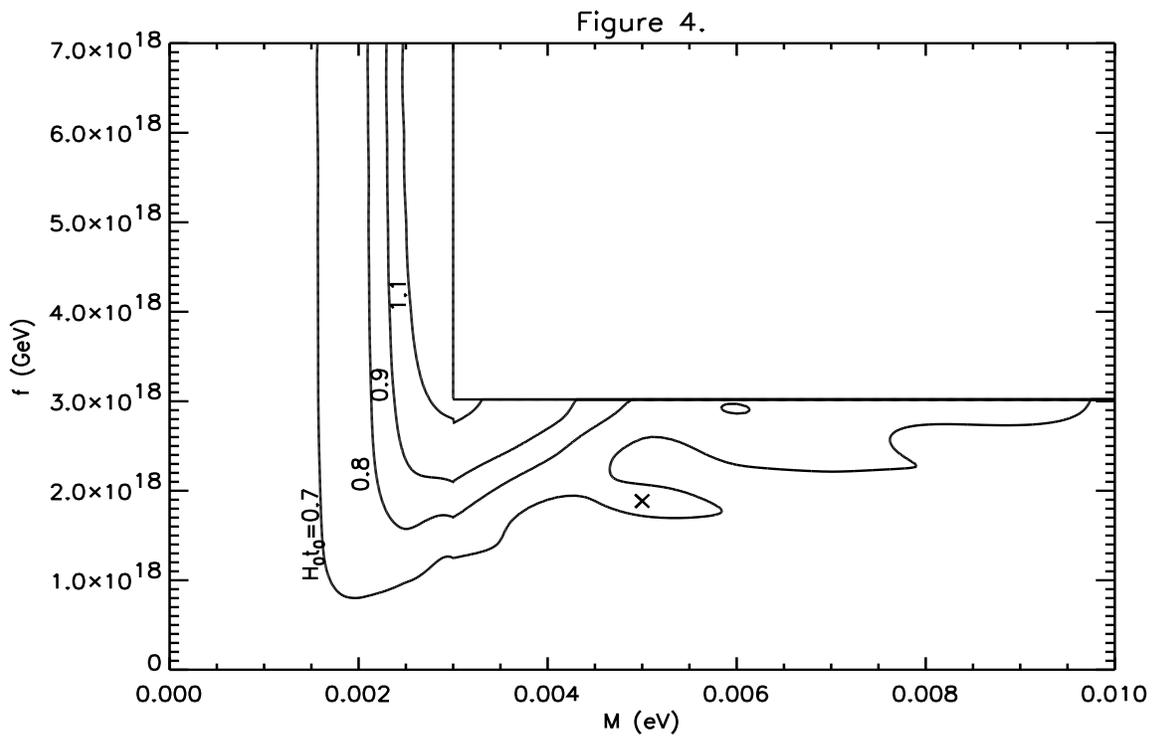}
}

\vspace{1. truecm}


\caption{Contours of $H_{0}t_{0}$ in the PNGB parameter
space for $\phi_i/f=1.6$. The cross indicates the same
model as in Fig. 3.}
\label{fig4}
\end{figure} 

\clearpage

\begin{figure}[t!]

\centering

\centerline{\epsfxsize=15. truecm \epsfysize=10. truecm 
\epsfbox{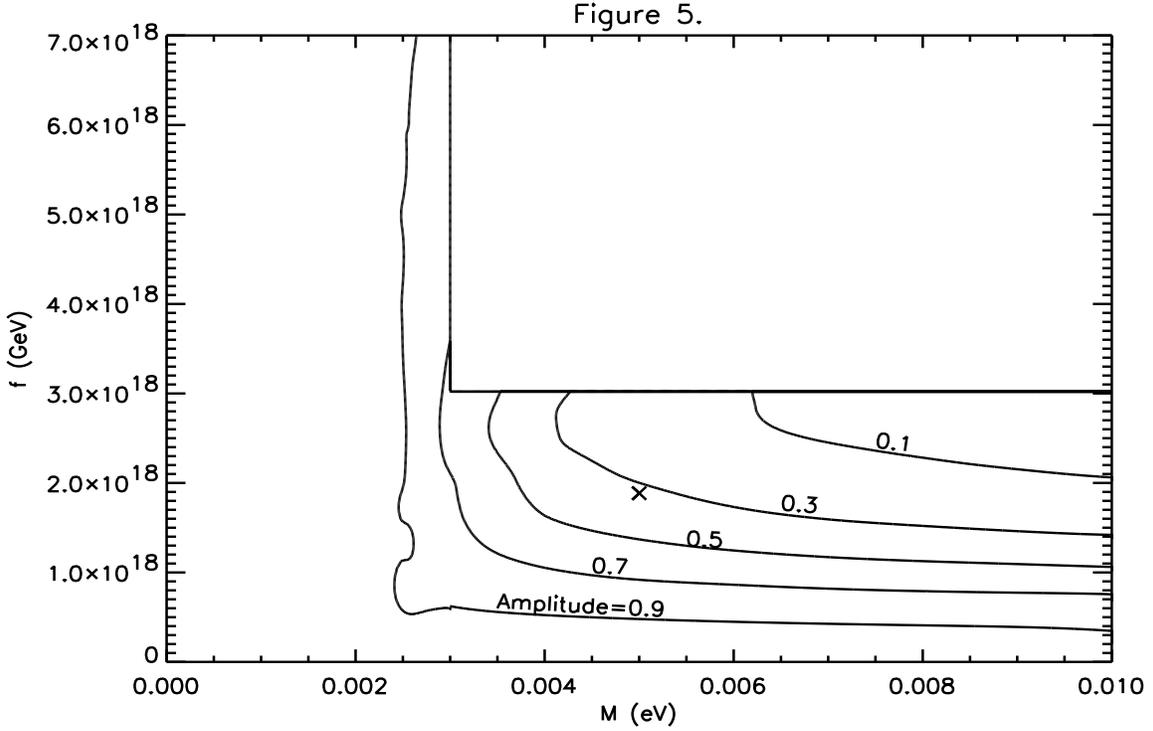}
}

\vspace{1. truecm}


\caption{Amplitude contours in the PNGB parameter
space for $\phi_i/f=1.6$. The amplitude shown is defined as
$\lim_{k \rightarrow 0} \left[{(P(k)/k)_\phi \over (P(k)/k)_\Lambda}
\right]$, the amplitude on large scales relative to that
of a $\Lambda$CDM model with
the same effective density as the PNGB
model, $\Omega_\Lambda = \Omega_\phi$.
Again, the cross marks the sample model for which
both the power spectrum shape and amplitude
provide a good fit to the galaxy clustering data.}
\label{fig5}
\end{figure}

\clearpage

\begin{figure}[t!]

\centering

\vspace{-2.0 truecm}

\centerline{\epsfxsize=15. truecm \epsfysize=15. truecm 
\epsfbox{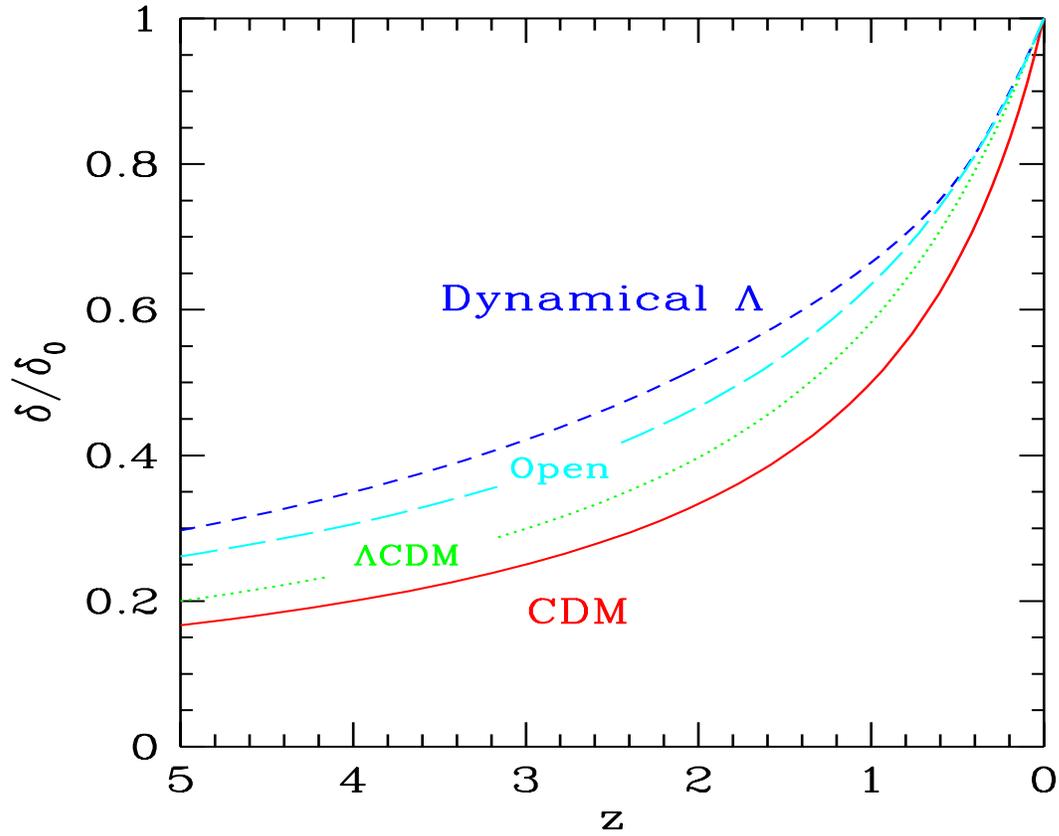}
}

\vspace{1. truecm}


\caption{Evolution of density perturbations. Shown is the density 
fluctuation amplitude at redshift $z$ normalized to its present 
amplitude, $\delta(z)/\delta_0$, vs. $z$. The models shown are 
standard $\Omega_m=1$ CDM (solid--red), $\Lambda$CDM with 
$\Omega_\Lambda = 0.6 =1-\Omega_m$ 
(dotted--green), an open CDM model with $\Omega_m=0.4$ (short dashed--light 
blue), and 
the dynamical $\Lambda$ model with $\Omega_\phi = 0.6$ (long dashed--dark blue
).}
\label{fig6}
\end{figure} 

\clearpage

\begin{figure}[t!]

\centering

\vspace{-1.5 truecm}

\centerline{\epsfxsize=15. truecm \epsfysize=15. truecm 
\epsfbox{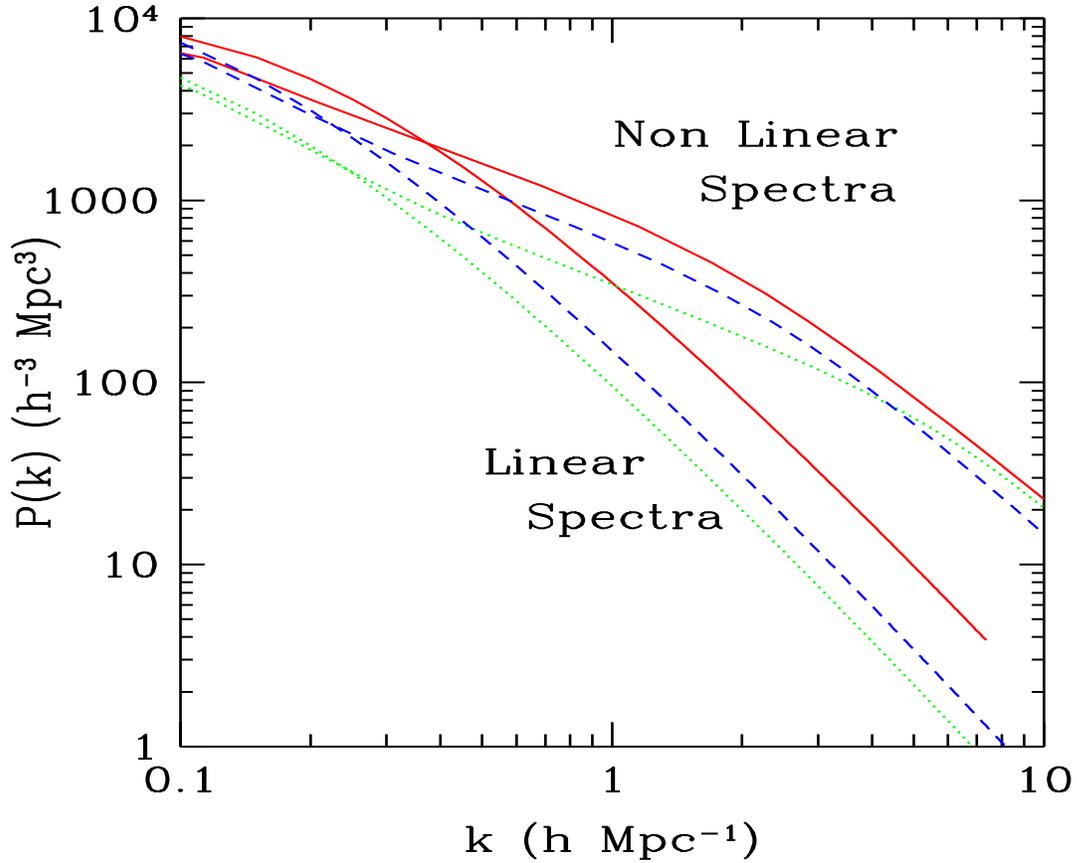}
}

\vspace{0.5 truecm}


\caption{Power spectra for COBE-normalized standard CDM (solid--red)  
with $\sigma_8=1.2$; 
$\Lambda$CDM with $\Omega_\Lambda = 0.6$ (dashed--blue), for which 
$\sigma_8=1.0$; 
and the dynamical $\Lambda$ model with $\Omega_\phi = 0.6$ (dotted--green), 
for 
which $\sigma_8=0.8$. The latter two models are normalied to the cluster 
abundance and have $h=0.7$.
Lower curves show the linear theory power spectra, upper 
curves the non-linear spectra obtained from scaling relations extracted 
from N-body simulations.}
\label{fig7}
\end{figure}

\begin{figure}

\centering

\centerline{\epsfxsize=15. truecm \epsfysize=10. truecm
\epsfbox{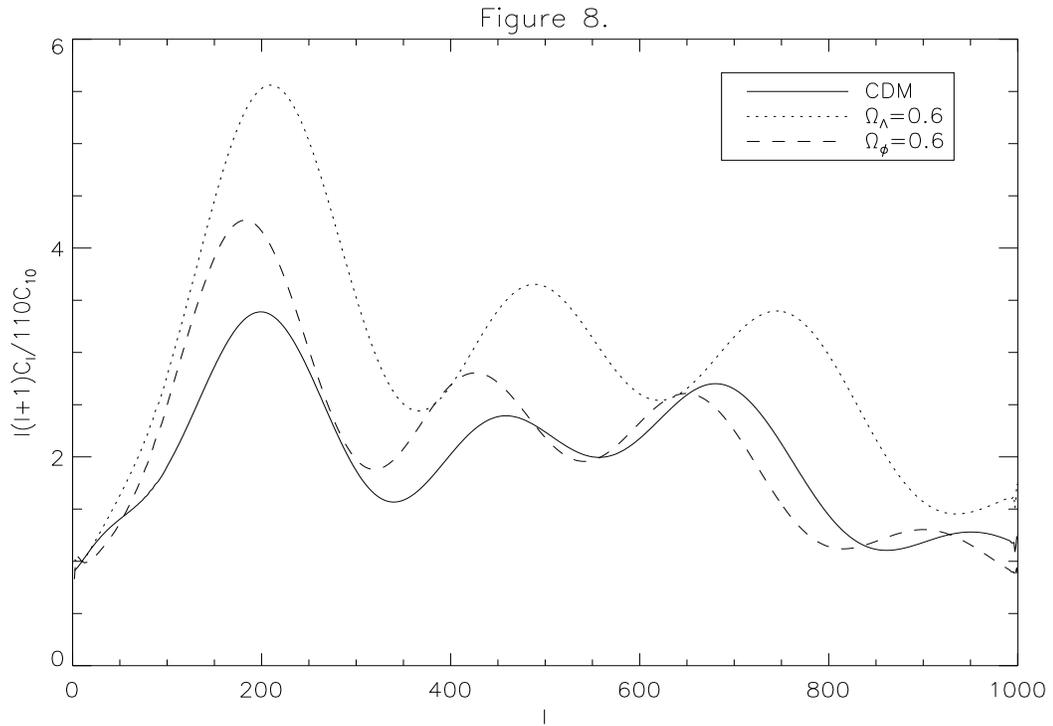}
}

\vspace{1. truecm}

\caption{CMBR angular power spectra for standard CDM,
$\Lambda$CDM with $\Omega_{\Lambda}=0.6$,
and scalar field $\Omega_{\phi}=0.6$ models.
In all models $h=0.7$, $\Omega_{B}=0.0255$, and $n=1$.
Plotted is $l(l+1)C_{l}$ vs. $l$, normalized at $l=10$.}
\label{fig8}
\end{figure}

\clearpage 

\begin{figure}

\centering

\centerline{\epsfxsize=15. truecm \epsfysize=10. truecm
\epsfbox{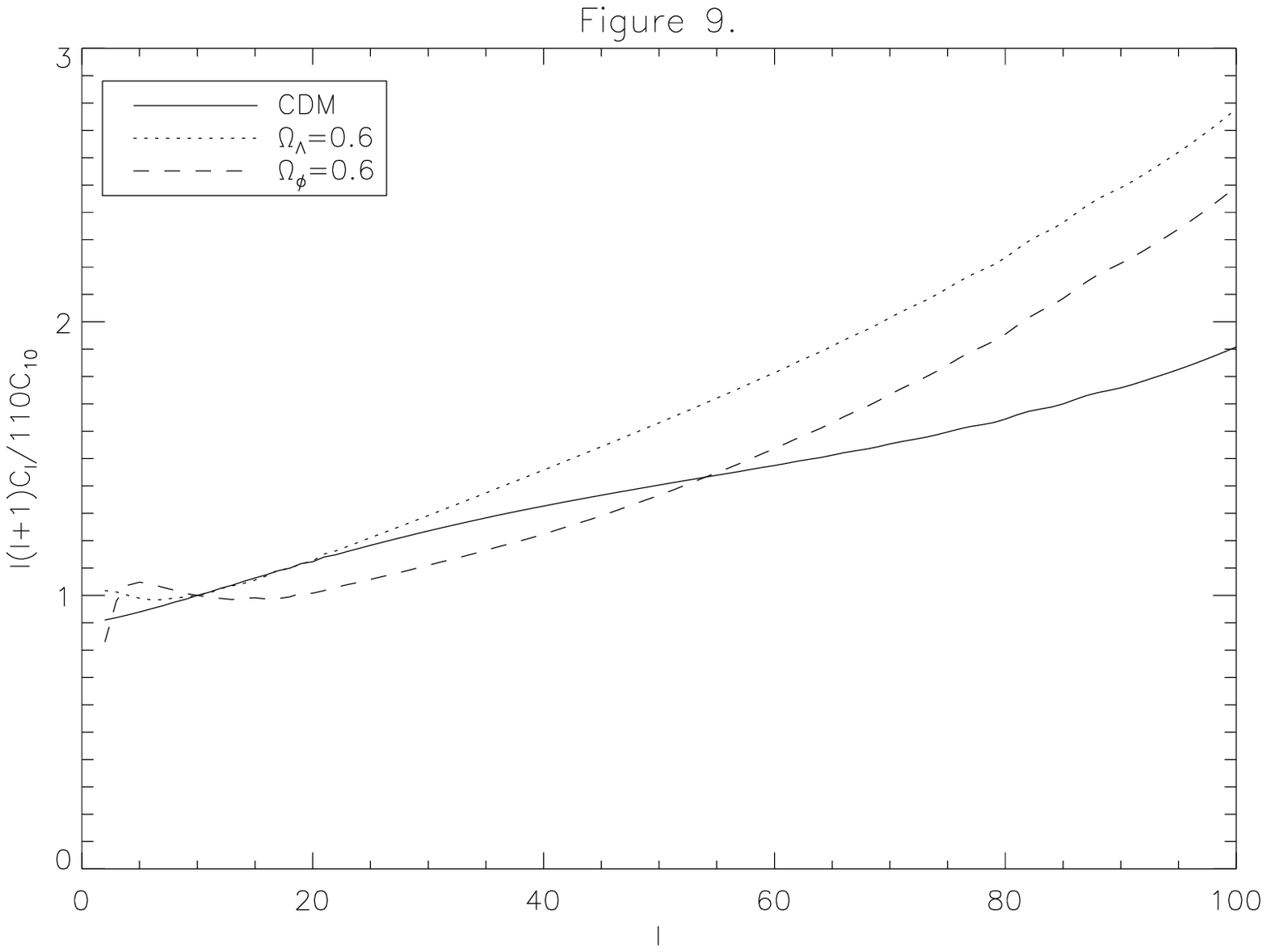}
}

\vspace{1. truecm}

\caption{Same as Fig. 8, but showing only the low
$l$ multipoles to emphasize the enhanced
ISW effect in the PNGB model.}
\label{fig9}
\end{figure}


\begin{thebibliography}{}
\bibitem{standard} Strictly speaking, ``standard'' CDM usually
refers to the parameter choice
$\Omega_m=1$ and $h=0.5$; while such a model yields
a better shape for $P(k)$ than an $\Omega_m=1$ model with $h=0.7$,
the fit to the data is still poor.
\end{thebibliography}
\end{document}